\documentclass[11pt,a4paper]{article}
\usepackage{fullpage}
\usepackage{epsfig}
\usepackage{amsfonts,amssymb,amstext,amsmath,amsthm}
\usepackage{shadow,times}
\usepackage{algorithm,algorithmic}
\usepackage{color}
\usepackage{xspace,paralist}

\allowdisplaybreaks

\newcommand{\sketch}[1]{}

\newtheorem{thm}{Theorem}[section]

\newtheorem{lemma}[thm]{Lemma}
\newtheorem{cor}[thm]{Corollary}

\newtheorem{claim}[thm]{Claim}

\def \RR   {{\mathbb R}}
\def \II   {{\mathbb I}}

\newcommand{\Prob}{{\rm{Pr}}}
\def\kcs {{\sf $k$-CS-PIP}\xspace}

\def\etal{{\em et al.}}
\def\sse{\subseteq}
\def\is{\mathcal{I}}

\def\floor#1{\lfloor {#1} \rfloor}

\newcommand{\set}[1]{\mathcal{#1}}
\newcommand{\ignore}[1]{}

\title{On $k$-Column Sparse Packing Programs}
\author{
Nikhil Bansal\thanks{IBM T.J. Watson Research Center, Yorktown
    Heights, NY 10598. email: \{nikhil,viswanath\}@us.ibm.com.}
\and Nitish Korula\thanks{Dept. of Computer Science, University of
    Illinois, Urbana IL 61801. Partially supported by NSF grant CCF
    07-28782 and a University of Illinois Dissertation Completion
    Fellowship. email: nkorula2@illinois.edu. }
\and Viswanath Nagarajan$^*$ \and Aravind Srinivasan\thanks{Dept.\ of Computer Science and
  Institute for Advanced Computer Studies, University of Maryland,
  College Park, MD 20742. Supported in part by NSF ITR Award
  CNS-0426683 and NSF Award CNS-0626636. email: srin@cs.umd.edu.}}

\date{}

\begin{document}

\maketitle

\begin{abstract}
  We consider the class of packing integer programs (PIPs) that are
  \emph{column sparse}, where there is a specified upper bound $k$ on
  the number of constraints that each variable appears in. We give an
  improved $(ek+o(k))$-approximation algorithm for $k$-column sparse PIPs.
  Our algorithm is based on a linear
  programming relaxation, and involves randomized rounding combined
  with alteration. We also show that the integrality gap of our LP
  relaxation is at least $2k-1$; it is known that even special cases
  of $k$-column sparse PIPs are $\Omega(\frac{k}{\log k})$-hard to
  approximate.

\smallskip

  We generalize our result to the case of maximizing
  monotone submodular functions over $k$-column sparse packing
  constraints, and obtain an $\smash{\left(\frac{e^2k}{e-1} + o(k)
    \right)}$-approximation algorithm. In obtaining this result, we
  prove a new property of submodular functions that generalizes the
  fractionally subadditive property,
  which  might be of independent interest.

\smallskip

  When the capacities of all constraints are large relative to the
  sizes, we obtain substantially better guarantees for these
  $k$-column sparse packing problems; again our result is tight (up to
  constant factors) relative to the natural LP relaxation.
\end{abstract}


\section{Introduction}
Packing integer programs (PIPs) are those of the form:
\begin{equation*}\label{eq:pip-defn} \max\left\{w^T\ x\mid Sx\le c,~x\in
    \{0,1\}^n\right\}, \quad \mbox{ where $w\in\RR^n_+,\, c\in
    \RR_+^m$ and } S\in \RR_+^{m\times n}.
\end{equation*} 
Above, $n$ is the number of variables/columns, $m$ is the number of
rows/constraints, $S$ is the matrix of {\em sizes}, $c$ is the {\em
  capacity} vector, and $w$ is the {\em weight} vector.
In general, PIPs are very hard to approximate: a special case is the
classic independent set problem, which is NP-Hard to approximate
within a factor of $n^{1-\epsilon}$ \cite{Z07}, whereas an
$n$-approximation is trivial.
Thus, various special cases of PIPs are often studied.  Here, we
consider {\em $k$-column sparse PIPs} (denoted \kcs), which are PIPs
where the number of non-zero entries in each column of matrix $S$ is
at most $k$. This is a fairly general class and models several basic
problems such as $k$-set packing \cite{HS89} and independent set in
graphs with degree at most $k$.

Recently, in a somewhat surprising result, Pritchard~\cite{P09} gave
an algorithm for \kcs where the approximation ratio only depends on
$k$; this is useful when $k$ is small. This result is surprising
because in contrast, no such guarantee is possible for $k$-row sparse
PIPs. In particular, the independent set problem on general graphs is
a 2-row sparse PIP, but is $n^{1-o(1)}$-hard to
approximate. Pritchard's algorithm \cite{P09} had an approximation
ratio of $2^k \cdot k^2$. Subsequently, an improved $O(k^2)$
approximation algorithm was obtained independently by Chekuri \etal
~\cite{CEK09b} and Chakrabarty-Pritchard~\cite{CP09}.

\paragraph{Our Results:} In this paper, we first consider the \kcs
problem and obtain an $(ek+o(k))$-approximation algorithm for it.  Our algorithm is based on solving a strengthened
version of the natural LP relaxation of \kcs, and then performing randomized rounding followed by suitable alterations.
In the {\em randomized rounding} step, we pick each variable independently (according to its LP value) and obtain a set
of variables with good expected weight; however, some constraints may be violated. Then in the \emph{alteration} step,
we drop some variables so as to satisfy all constraints, while still having good expected weight. A similar approach
can be used with the natural relaxation for \kcs obtained by simply dropping the integrality constraints on the
variables; this gives a slightly weaker $8k$-approximation bound. However, the analysis of this weaker result is much
simpler and we thus present it first.  To obtain the $ek + o(k)$ bound, we construct a stronger LP relaxation by adding
additional valid constraints to the natural relaxation for \kcs. The analysis of our rounding procedure is based on
exploiting these additional constraints and using the positive correlation between various probabilistic events via the
FKG inequality.

Our result is almost the best possible that one can hope for using the LP based approach.  We show that the integrality
gap of the strengthened LP is at least $2k-1$, so our analysis is tight up to a small constant factor $e/2 \approx
1.36$ for large values of $k$. Even without restricting to LP based approaches, an $O(k)$ approximation is nearly best
possible since it is NP-Hard to obtain an $o(k/\log k)$-approximation for the special case of $k$-set packing
\cite{HSS06}.  We also obtain improved results for \kcs when capacities are large relative to the sizes.
In particular, we obtain a $\Theta(k^{1/\floor{B}})$-approximation algorithm for \kcs, where $B:=\min_{i\in[n],
j\in[m]} \, c_j/s_{ij}$ measures the relative slack between the capacities $c$ and sizes $S$. We also show that this
result is tight up to constant factors relative to its LP relaxation.

Our second main result is for the more general problem of maximizing a monotone submodular function over packing
constraints that are $k$-column sparse. This problem is a common generalization of maximizing a submodular function
over (a) a $k$-dimensional knapsack~\cite{KST09}, and (b) the intersection of $k$ partition matroids~\cite{NWF78II}.
Here, we obtain an $\left(\frac{e^2k}{e-1} +
  o(k) \right)$-approximation algorithm for this problem. Our
algorithm uses the continuous greedy algorithm of Vondr{\'a}k~\cite{V08} in conjunction with our randomized rounding
plus alteration based approach.  However, it turns out that the analysis of the approximation guarantee is much more
intricate: In particular, we need a generalization of a result of Feige \cite{F06} that shows that submodular functions
are also \emph{fractionally
  subadditive}. See Section~\ref{sec:submod} for a statement of the
new result, Theorem~\ref{thm:subadd}, and related context.  This generalization is based on an interesting connection
between submodular functions and the FKG inequality.  We believe that this result and technique might be of further use
in the study of submodular optimization.


\paragraph{Related Previous Work:}
Various special cases of \kcs have been extensively studied. An important special case is the {\em $k$-set packing}
problem, where given a collection of sets of cardinality at most $k$, the goal is to find the maximum weight
sub-collection of mutually disjoint sets. This is equivalent to \kcs where the constraint matrix $S$ is 0-1 and the
capacity $c$ is all ones. Note that for $k=2$ this is {\em maximum
  weight matching} which can be solved in polynomial time, and for
$k=3$ the problem becomes APX-hard~\cite{HSS06}. After a long line of work \cite{HS89,AH97,CH99,B00}, the best-known
approximation ratio for this problem is $\frac{k+1}2+\epsilon$ obtained using local search techniques~\cite{B00}.  An
improved bound of $\frac{k}2+\epsilon$ is also known~\cite{HS89} for the unweighted case, i.e., the weight vector
$w=\mathbf{1}$. It is also known that the natural LP relaxation for this problem has integrality gap at least
$k-1+1/k$, and in particular this holds for the projective plane instance of order $k-1$. Hazan \etal~\cite{HSS06}
showed that $k$-set packing is $\Omega(\frac{k}{\log k})$-hard to approximate.

Another special case of \kcs is the independent set problem in graphs with maximum degree at most $k$.  This is
equivalent to \kcs where the constraint matrix $S$ is 0-1, capacity $c$ is all ones, and each row is $2$-sparse. This
problem has an $O(k \log \log k/\log k)$-approximation \cite{H02}, and is $\Omega(k/\log^2 k)$-hard to approximate
\cite{AKS09}, assuming the Unique Games Conjecture \cite{Khot2}.

Shepherd and Vetta~\cite{SV07} studied the {\em demand matching} problem on graphs, which is \kcs with $k=2$, with the
further restriction that in each column the non-zero entries are equal, and that no two columns have non-zero entries
in the same two rows.
They gave an LP-based $3.264$-approximation algorithm \cite{SV07}, and showed that the natural LP relaxation for this
problem has integrality gap at least $3$. They also showed the demand matching problem to be APX-hard even on bipartite
graphs. For larger values of $k$, problems similar to {\em demand matching} have been studied under the name of
column-restricted PIPs \cite{KS}, which arise in the context of routing flow unsplittably (see also
\cite{BavejaSa,BavejaSb}).  In particular, an $11.54 k$-approximation algorithm was known \cite{CMS07} where (i) in
each column all non-zero entries are equal, and (ii) the maximum entry in $S$ is at most the minimum entry in $c$ (this
is also known as the no bottle-neck assumption); later, it was observed in \cite{CEK09a} that even without the second
of these conditions, one can obtain an $8k$ approximation. The literature on unsplittable flow is quite extensive; we
refer the reader to \cite{BFKS,CEK09a} and references therein.


For the general \kcs, Pritchard~\cite{P09} gave a $2^kk^2$-approximation algorithm, which was the first result with
approximation ratio depending only on $k$. Pritchard's algorithm was based on solving an iterated LP relaxation, and
then applying a randomized selection procedure. Independently, \cite{CEK09b} and \cite{CP09} showed that this final
step could be derandomized, yielding an improved bound of $O(k^2)$. All these previous results crucially use the
structural properties of basic feasible solutions of the LP relaxation. However, as stated above, our result is based
on randomized rounding with alterations and does not use properties of basic solutions. This is crucial for the
submodular maximization version of the problem, as a solution to the fractional relaxation there does not have these
properties.

We remark that randomized rounding with alteration has also been used earlier by Srinivasan~\cite{S01} in the context
of PIPs. However, the focus of this paper is different from ours; in previous work \cite{S99}, Srinivasan had bounded
the integrality gap for PIPs by showing a randomized algorithm that obtained a ``good'' solution (one that satisfies
all constraints) with positive --- but perhaps exponentially small --- probability. In \cite{S01}, he proved that
rounding followed by alteration leads to an efficient and parallelizable algorithm; the rounding gives a ``solution''
of good value in which most constraints are satisfied, and one can alter this solution to ensure that all constraints
are satisfied. (We note that \cite{S99,S01} also gave derandomized versions of these algorithms.)

Related issues have been considered in discrepancy theory, where the goal is to round a fractional solution to a
$k$-column sparse linear program so that the capacity violation for any constraint is minimized. A celebrated result of
Beck-Fiala \cite{BF} shows that the capacity violation is at most $O(k)$.  A major open question in discrepancy theory
is whether the above bound can be improved to $O(\sqrt{k})$, or even $O(k^{1-\epsilon})$ for some $\epsilon>0$. While
the result of \cite{P09} uses techniques similar to that of \cite{BF}, a crucial difference in our problem is that no
constraint can be violated at all.  In fact, at the end of Section \ref{sec:kcs}, we show another crucial qualitative
difference between discrepancy and \kcs.

There is a large body of work on constrained maximization of submodular functions; we only cite the relevant papers
here. Calinescu \etal~\cite{CCPV07} introduced a continuous relaxation (called the \emph{multi-linear extension} or
extension-by-expectation) of submodular functions and subsequently Vondr{\'a}k~\cite{V08} gave an elegant
$\frac{e}{e-1}$-approximation algorithm for solving this continuous relaxation over any ``downward monotone" polytope
$\mathcal{P}$, as long as there is a polynomial-time algorithm for optimizing linear functions over $\mathcal{P}$. We
use this continuous relaxation in our algorithm for submodular maximization over $k$-sparse packing constraints. As
noted earlier, $k$-sparse packing constraints generalize both $k$-partition matroids and $k$-dimensional knapsacks.
Nemhauser \etal~\cite{NWF78II} gave a $(k+1)$-approximation for submodular maximization over the intersection of $k$
partition matroids; when $k$ is constant, Lee \etal~\cite{LMNS09} improved this to $k+\epsilon$. Kulik
\etal~\cite{KST09} gave an $\left(\frac{e}{e-1}+\epsilon\right)$-approximation for submodular maximization over
$k$-dimensional knapsacks when $k$ is constant; if $k$ is part of the input, the best known approximation bound is
$O(k)$.

\paragraph{Problem Definition and Notation:}
Before we begin, we formally describe the \kcs problem and fix some notation. Let the items (i.e., columns) be indexed
by $i\in[n]$ and the constraints (i.e., rows) be indexed by $j\in[m]$. We consider the following packing integer
program.
\begin{equation*}
\max \left\{ \sum_{i =1 }^n w_i x_i  \,\, \big| \,\, \sum_{i=1}^n s_{ij}\cdot x_i \le c_j,\, \forall \,j\in [m]; \,\,\,
x_i \in \{0,1\}, \, \forall\, i\in[n]\right\}
\end{equation*}
\ignore{\vspace{-3mm}
\begin{align}
\max \quad & \sum_{i =1 }^n w_i x_i \notag\\
\mbox{s.t.} \quad & \sum_{i=1}^n s_{ij}\cdot x_i \le c_j, & & \forall j\in [m] \notag \\
& x_i \in \{0,1\}, & & \forall i\in[n]. \notag
\end{align}}

We say that item $i$ \emph{participates} in constraint $j$ if $s_{ij}
> 0$. For each $i\in[n]$, let $N(i) :=\{j\in [m]\mid s_{ij}>0\}$ be
the set of constraints that $i$ participates in. In a $k$-column sparse PIP, we have $|N(i)| \leq k$ for each
$i\in[n]$.  The goal is to find the maximum weight subset of items such that all the constraints are satisfied.

We define the {\em slack} as $B:=\min_{i\in[n], j\in[m]} \, c_j/s_{ij}$. By scaling the constraint matrix, we may
assume that $c_j=1$ for all $j\in[m]$. We also assume that $s_{ij} \leq 1$ for each $i,j$; otherwise, we can just fix
$x_i=0$. Finally, for each constraint $j$, we let $P(j)$ denote the set of items participating in this constraint. Note
that $|P(j)|$ can be arbitrarily large.

\paragraph{Organization:}
In Section~\ref{sec:kcs} we begin with the natural LP relaxation, and
describe a simple algorithm with approximation ratio $8k$.
We then present a stronger relaxation, and use it to obtain an $(e+o(1))k$-approximation.  We also present the
integrality gap of $2k-1$ for this strengthened LP, implying that our result is almost tight. In
Section~\ref{sec:submod}, we describe the $\left(\frac{e^2}{e-1}+o(1)\right)k$-approximation for $k$-column sparse
packing problems over a submodular objective. Finally, in Section~\ref{app:B}, we deal with the \kcs problem when the
capacities of all constraints are large relative to the sizes, and obtain  significantly better approximation ratios.
Again there is a matching integrality gap up to a constant factor.

\section{Approximation Algorithms for \kcs}\label{sec:kcs}
Before presenting our algorithm, we describe a (seemingly correct)
algorithm that does not quite work. Understanding why this easier
algorithm fails gives useful insight into the design for the correct
algorithm.

\paragraph{A strawman Algorithm:} Consider the following
algorithm. Let $x$ be some optimum solution to the natural LP
relaxation of \kcs (i.e. dropping integrality).  For each element $i
\in [n]$, select it independently at random with probability $
x_i/(2k)$. Let $\set{S}$ be the chosen set of items. For any
constraint $j \in [m]$, if it is violated, then discard all items
$\set{S}\cap P(j)$, i.e. items $i\in\set{S}$ for which $s_{ij} >0$.

Since the probabilities are scaled down by $2k$, by Markov's
inequality any constraint $j$ is violated with probability at most
$1/(2k)$. Hence, any constraint will discard its items with
probability at most $1/2k$. By the $k$-sparse property, each element
can be discarded by at most $k$ constraints, and hence by union bound
over those $k$ constraints, it is discarded with probability at most
$k \cdot (1/2k) = 1/2$. Since an element is chosen in $\set{S}$ with
probability $x_i/2k$, this implies that it lies in the overall
solution with probability at least $x_i/(4k)$, implying that the
proposed algorithm is a $4k$ approximation.

However, the above argument is not correct. Consider the following
example. Suppose there is a single constraint (and so $k=1$),
$$ M x_1 + x_2 + x_3 + x_4 + \ldots + x_M \leq M $$
where $M \gg 1$ is a large integer. Clearly, setting $x_i=1/2$ for
$i=1,\ldots,M$ is a feasible solution. Now consider the execution of
the strawman algorithm. Note that whenever item 1 is chosen in
$\set{S}$, it is very likely that some item other than 1 will also be
chosen (since $M \gg 1$ and we pick each item independently with
probability $x_i/2k= 1/4$); in this case, item 1 would be
discarded. Thus the final solution will almost always {\em not
  contain} item 1, violating the claim that it lies in the final
solution with probability at least $x_1/4k=1/8$.

The key point is that we must consider the probability of an item
being discarded by some constraint, {\em conditional} on it being
chosen in the set $\set{S}$ (for item 1 in the above example, this
probability is close to one, not at most half). This is not a problem
if either all item sizes are small (i.e. say $s_{ij} \leq c_j/2$), or
all item sizes are large (say $s_{ij} \approx c_j$). The algorithm we
analyze shows that the difficult case is indeed when some constraints
contain both large and small items, as in the example above.

\ignore{This is tight up to a very small constant factor, less than
  $1.36$, for large values of $k$.  The improvement comes from
  \emph{four} different steps. First, we use we use a slightly
  stronger LP relaxation. Second, we modify the alteration/deletion
  step to obtain $\set{S}'$ from $\set{S}$; the previous algorithm
  sometimes deleted items even when constraints were not violated, or
  deleted more items than necessary. Third, we are more careful in
  analyzing the probability that constraint $j$ causes item $i$ to be
  deleted from the set $\set{S}$; we take into account a certain
  discreteness of item sizes.  And fourth, for each item $i$, we use
  the FKG inequality instead of the weaker union bound over all
  constraints in $N(i)$.}

\subsection{A Simple Algorithm for \kcs} \label{app:simple}
In this subsection, we use the obvious LP relaxation for \kcs (i.e. dropping the integrality condition) and obtain an
$8k$-approximation algorithm. An item $i\in[n]$ is called {\em big} for constraint $j\in[m]$ iff $s_{ij}>\frac12$; and
$i$ is {\em small} for constraint $j$ iff $0<s_{ij}\le \frac12$. The algorithm first solves the LP relaxation to obtain
an optimal fractional solution $x$. Then we round to an integral solution as follows. With foresight, set $\alpha=4$.

\begin{enumerate}
\item Sample each item $i\in[n]$ independently with probability
  $x_i/(\alpha k)$. \\Let $\set{S}$ denote the set of chosen items.
  We call an item in $\set{S}$ an $\set{S}$-item.

\item For each item $i$, mark $i$ (for deletion) if, for any constraint
  $j \in N(i)$, either:
  \begin{itemize}
  \item $\set{S}$ contains some \emph{other} item $i'\in[n]\setminus \{i\}$ which is big
    for constraint $j$ or

  \item The sum of sizes of $\set{S}$-items that are small for $j$
    exceeds $1$.  (i.e. the capacity).
  \end{itemize}

\item Delete all marked items, and return $\set{S}'$, the set of remaining items.

\end{enumerate}

\paragraph{Analysis:}
We will show that this algorithm gives an $8k$ approximation.

\begin{lemma}\label{lem:feas}
  Solution $\set{S}'$ is feasible with probability one.
\end{lemma}
\begin{proof} Consider any fixed constraint $j\in[m]$.
\begin{enumerate}
\item Suppose there is some $i' \in \set{S}'$ that is big for
  $j$. Then the algorithm guarantees that $i'$ will be the only item
  in $\set{S}'$ (either small or big) that participates in constraint
  $j$: Consider any other $\set{S}$-item $i$ participating in $j$; $i$
  must have been deleted from $\set{S}$ because $\set{S}$ contains
  another item (namely $i'$) that is big for constraint $j$.  Thus,
  $i'$ is the only item in $\set{S}'$ participating in constraint $j$,
  and so the constraint is trivially satisfied, as all sizes $\leq
  1$.

\item The other case is when all items in $\set{S}'$ are small for
  $j$. Let $i\in \set{S}'$ be some item that is small for $j$ (if
  there are none such, then constraint $j$ is trivially satisfied).
  Since $i$ was not deleted from $\set{S}$, it must be that the total
  size of $\set{S}$-items that are small for $j$ did not
  exceed  $1$. Now, $\set{S}' \subseteq \set{S}$, and so this
  condition is also true for items in $\set{S}'$.
\end{enumerate}
Thus every constraint is satisfied by solution $\set{S}'$ and we obtain the lemma.
\end{proof}

We now prove the main theorem.

\begin{thm} \label{thm:main}
  For any item $i\in[n]$, the probability $\Prob[i\in \set{S}'\mid
  i\in \set{S}]\ge 1-\frac2\alpha$. Equivalently, the probability that
  item $i$ is deleted from $\set{S}$ conditional on it being chosen in
  $\set{S}$ is at most $2/\alpha$.
\end{thm}
\begin{proof}
  For any item $i$ and constraint $j \in N(i)$, let $B_{ij}$ denote
  the event that $i$ is marked for deletion from $\set{S}$ because
  there is some other $\set{S}$-item that is big for constraint
  $j$. Let $G_j$ denote the event that the total size of
  $\set{S}$-items that are small for constraint $j$ exceeds $1$.  For
  any item $i\in[n]$ and constraint $j\in N(i)$, we will show that:
  \begin{equation}  \label{eq:big-small}
    \Prob[B_{ij}\mid i\in \set{S}] + \Prob[G_j\mid i\in \set{S}]
    \le \frac2{\alpha k}
  \end{equation}

  We prove~\eqref{eq:big-small} using the following intuition: The
  total extent to which the LP selects items that are \emph{big} for
  any constraint cannot be more than $2$ (each big item has size at
  least $1/2$); therefore, $B_{ij}$ is unlikely to occur since we
  scaled down probabilities by factor $\alpha k$. Ignoring for a
  moment the conditioning on $i \in \set{S}$, event $G_j$ is also
  unlikely, by Markov's Inequality. But items are selected for
  $\set{S}$ independently, so if $i$ is big for constraint $j$, then
  its presence in $\set{S}$ does not affect the event $G_j$ at all. If
  $i$ is small for constraint $j$, then \emph{even if $i \in
    \set{S}$}, the total size of $\set{S}$-items is unlikely to exceed
  $1$.

  \medskip
  We now prove~\eqref{eq:big-small} formally, using some care to save
  a factor of $2$.  Let $B(j)$ denote the set of items that are big
  for constraint $j$, and $Y_{j} := \sum_{\ell\in B(j)} x_\ell$. By
  the LP constraint for $j$, it follows that $Y_j\le 2$ (since each
  $\ell \in B(j)$ has size $s_{\ell j}>\frac12$). Now by a union
  bound,
  \begin{equation}\label{eq:big-blk}
    \Prob[B_{ij}\mid i\in \set{S}] \le \frac1{\alpha k}
    \sum_{\ell \in B(j)\setminus \{i\}} x_\ell \le \frac{Y_j}{\alpha k}
    \le \frac2{\alpha k}.
  \end{equation}

  Now, let $G_{-i}(j)$ denote the set of items that are small for
  constraint $j$, \emph{not counting} item $i$, even if it is
  small. Using the LP constraint $j$, we have:
  \begin{equation}\label{eq:exp-small}
    \sum_{\ell\in G_{-i}(j)} s_{\ell j}\cdot x_l \le 1 - \sum_{\ell\in
      B(j)} s_{\ell j}\cdot x_\ell \le 1-\frac{Y_j}{2}.
  \end{equation}

  Since each item $i'$ is chosen into $\set{S}$ with probability
  $x_{i'}/(\alpha k)$, inequality~\eqref{eq:exp-small} implies that
  the expected total size of $\set{S}$-items in $G_{-i}(j)$ is at most
  $\frac1{\alpha k}\left( 1- Y_j/2\right)$. By Markov's inequality,
  the probability that the total size of these $\set{S}$-items exceeds
  $1/2$ is at most $\frac2{\alpha k}\left( 1- Y_j/2\right)$. Since
  items are chosen independently and $i \not \in G_{-i}(j)$, we obtain
  this probability even conditioned on $i \in \set{S}$.

  If $i$ is big for $j$, event $G_j$ occurs only if the total size of
  $\set{S}$-items in $G_{-i}(j)$ exceeds $1$. If $i$ is small for $j$,
  event $G_j$ occurs only if the total size of small $\set{S}$-items
  participating in $j$ exceeds $1$; as $s_{ij} \le 1/2$, the total
  size of $\set{S}$-items in $G_{-i}(j)$ must exceed $1/2$. Thus,
  whether $i$ is big or small,
  $$\Prob[G_j \mid i\in \set{S}] \le \frac2{\alpha k}
  \left( 1- \frac{Y_j}2\right) = \frac2{\alpha k} - \frac{Y_j}{\alpha k}.$$

  Combined with inequality~\eqref{eq:big-blk} we obtain
  \eqref{eq:big-small}:
  $$\Prob[B_{ij}\mid i\in \set{S}] + \Prob[G_{j}\mid i\in \set{S}]
  \le \frac{Y_j}{\alpha k} + \Prob[G_j \mid i\in \set{S}]
  \le \frac{Y_j}{\alpha k} + \frac2{\alpha k} - \frac{Y_j}{\alpha k}
  = \frac2{\alpha k}.$$

  To see that~\eqref{eq:big-small} implies the theorem, for any item
  $i$, simply take the union bound over all $j \in N(i)$. Thus, the
  probability that $i$ is deleted from $\set{S}$ conditional on it
  being chosen in $\set{S}$ is at most $2/\alpha$. Equivalently,
  $\Prob[i \in \set{S'} \mid i \in \set{S}] \ge 1 - 2/\alpha$.
\end{proof}

We are now ready to prove the final result.
\begin{thm}\label{thm:simple-result}
  There is a randomized $8k$-approximation algorithm for \kcs.
\end{thm}
\begin{proof}
  First observe that our algorithm always outputs a feasible solution
  (Lemma~\ref{lem:feas}). To bound the objective value, recall that
  $\Prob[i\in \set{S}]=\frac{x_i}{\alpha k}$ for all $i\in[n]$. Hence
  Theorem~\ref{thm:main} implies that

$$\Prob[i\in \set{S}'] \ge \Prob[ i \in \set{S}]
   \cdot \Prob[i\in \set{S}'| i \in \set{S}] \ge
   \frac{x_i}{\alpha k} \cdot \left( 1 - \frac{2}{\alpha} \right)$$

   for all $i\in[n]$. Finally using linearity of expectation and
   $\alpha=4$, we obtain the theorem.
\end{proof}

\noindent {\em Remark:} We note that the analysis above only uses Markov's inequality conditioned on a single item
being chosen in set $\set{S}$. Thus a pairwise independent distribution suffices to choose the set $\set{S}$, and hence
the algorithm can be easily derandomized.\\

\noindent {\em General upper bounds:} The \kcs problem as defined assumes all variables to be 0-1. We note that our
result easily extends to the \kcs problem with general upper bounds on variables. Assuming an LP-based
$\rho$-approximation algorithm for \kcs with unit upper-bounds, it is straightforward to obtain a
$(\rho+1)$-approximation for \kcs with general upper-bounds. The algorithm first solves the natural LP relaxation to
obtain fractional solution $y\in \mathbb{R}_+^n$. Let $z\in\mathbb{Z}_+^n$ and $x\in [0,1]^n$ be defined as:
$z_i=\lfloor y_i\rfloor$ and $x_i = y_i - \lfloor y_i\rfloor$ for all $i\in [n]$; note that $w^T\,y = w^T\,z + w^T\,x$.
Clearly $z$ is a feasible integral solution. Moreover $x$ is a feasible fractional solution to the same \kcs instance
even with {\em unit upper-bounds}. Hence using the rounding algorithm of this subsection, we obtain a feasible integral
solution $\overline{x}\in \{0,1\}^n$ with $w^T \, \overline{x} \ge \frac1\rho\cdot w^T\,x$. It can be seen by simple
calculation that the better of $z$ and $\overline{x}$ is a $(\rho+1)$-approximate solution relative to the natural LP
relaxation for \kcs with general upper-bounds.

\subsection{A Stronger LP,  and Improved Approximation}
We now present our strengthened LP and the $(ek+o(k))$-approximation algorithm for \kcs.



\paragraph{ Stronger LP relaxation.} Recall that entries are scaled so
that all capacities are one. An item $i$ is called {\em big} for
constraint $j$ iff $s_{ij} > 1/2$. For each constraint $j\in [m]$, let
$B(j) = \{i\in [n]\mid s_{ij}>\frac12\}$ denote the set of big
items. Since no two items that are big for some constraint can be
chosen in an integral solution, the inequality $\sum_{i\in B(j)} x_i
\le 1$ is valid for each $j\in [m]$. The strengthened LP relaxation
that we consider is as follows.
\begin{align}
\max \quad & \sum_{i =1 }^n w_i x_i  \label{lp:first} \\
\mbox{s.t.} \quad & \sum_{i=1}^n s_{ij}\cdot x_i \le c_j, & & \forall j\in [m] \\
&   \sum_{i\in B(j)} x_i \le 1,  & &  \forall j\in [m]. \label{eq:new-constraint}\\
 &  0 \le x_i \le 1, & & \forall i\in[n]. \label{lp:last}
\end{align}

\noindent  {\bf Algorithm:} The algorithm obtains an optimal solution $x$ to the LP
relaxation~(\ref{lp:first}-\ref{lp:last}), and rounds it to an integral solution $\set{S}'$ as follows (parameter
$\alpha$ will be set to $1$ later).
\begin{enumerate}
\item Pick each item $i\in[n]$ independently with probability
  $x_i/(\alpha k)$. Let $\set{S}$ denote the set of chosen items.
\item For any item $i$ and constraint $j \in N(i)$, let $E_{ij}$
  denote the event that the items $\{i'\in \set{S} \mid s_{i'j}\ge
  s_{ij}\}$ have total size (in constraint $j$) exceeding one. Mark
  $i$ for deletion if $E_{ij}$ occurs for any $j \in N(i)$.
\item Return set $\set{S}'\sse \set{S}$ consisting of all items $i\in
  \set{S}$ not marked for deletion.
\end{enumerate}

Note the rule for deleting an item from $\set{S}$. In particular, whether item $i$ is deleted from constraint $j$ only
depends on items that are at least as large as $i$ in $j$.



\paragraph{ Analysis:} It is clear that $\set{S}'$ is feasible with probability one. The main lemma is the
following, where we show that each item appears in $\set{S}'$ with good probability.

\begin{lemma}\label{lem:prob-eij}
For every item $i\in[n]$ and constraint $j\in N(i)$, we have  $\Prob[E_{ij} \mid i \in \set{S}] \le \frac{1}{\alpha k}
\left(1 + (\frac{2}{\alpha k})^{1/3} \right)$.
\end{lemma}
\begin{proof}
  Let $\ell:= (4\alpha k)^{1/3}$. We classify items in relation to
  constraints as:
  \begin{itemize}
    \item Item $i\in [n]$ is {\em big} for constraint $j\in[m]$ if
      $s_{ij}>\frac12$.
    \item Item $i\in [n]$ is {\em medium} for constraint $j\in[m]$ if
      $\frac1\ell \le s_{ij}\le \frac12$.
    \item Item $i\in [n]$ is {\em tiny} for constraint $j\in[m]$ if
      $s_{ij}< \frac1\ell$.
  \end{itemize}

  For any constraint $j\in[m]$, let $B(j), M(j), T(j)$ respectively
  denote the set of big, medium, tiny items for $j$.
  In the next three claims, we bound $\Prob[E_{ij} | i \in \set{S}]$
  when item $i$ is big, medium, and tiny respectively.

  \begin{claim}\label{cl:big-bnd}
    For any $i\in [n]$ and $j\in [m]$ s.t. item $i$ is
    big for constraint $j$, $\Prob[E_{ij}\mid i\in \set{S}] \le
    \frac1{\alpha k}$.
  \end{claim}
  \begin{proof}
    The event $E_{ij}$ occurs if some item that is at least as large
    as $i$ for constraint $j$ is chosen in $\set{S}$.  Since $i$ is
    big in constraint $j$, $E_{ij}$ occurs only if some big item other
    than $i$ is chosen for $S$. Now by the union bound, the
    probability that some item from $B(j)\setminus \{i\}$ is chosen
    into $\set{S}$ is:
    $$\Prob\left[\left( B(j)\setminus \{i\}\right) \bigcap \set{S} \ne
      \emptyset ~\big|~ i\in \set{S} \right]
    \le \sum_{i'\in B(j) \setminus \{i\}} \frac{x_{i'}}{\alpha k}
    \le \frac1{\alpha k}\sum_{i'\in B(j)} x_{i'}
    \le \frac1{\alpha k},$$
    where the last inequality follows from the new LP
    constraint~\eqref{eq:new-constraint} on big items for $j$.
  \end{proof}

  \begin{claim}\label{cl:med-bnd}
    For any $i\in [n]$, $j\in [m]$ s.t. item $i$ is medium for
    constraint $j$, $\Prob[E_{ij}\mid i\in \set{S}] \le \frac1{\alpha k}
    \left(1+ \frac{\ell^2}{2\alpha k}\right)$.
  \end{claim}
  \begin{proof}
    Here, if event $E_{ij}$ occurs then it must be that either some
    big item is chosen or (otherwise) at least two medium items other
    than $i$ are chosen, i.e. $E_{ij}$ implies that either
    $\set{S}\bigcap B(j)\ne \emptyset$ or $|\set{S}\bigcap
    \left(M(j)\setminus \{i\}\right)|\ge 2$.  This is because $i$
    together with any \emph{one} other medium item is not enough to
    reach the capacity of constraint $j$.  (Since $i$ is medium, we do
    not consider tiny items for constraint $j$ in determining whether
    $i$ should be deleted.)

    Just as in Claim~\ref{cl:big-bnd}, we have that the probability
    some big item for $j$ is chosen is at most $1/\alpha k$, i.e.
    $\Prob\left[\set{S}\bigcap B(j)\ne \emptyset \mid i\in
      \set{S}\right] \le \frac1{\alpha k}$.

    Now consider the probability that $|\set{S}\bigcap
    \left(M(j)\setminus \{i\}\right)|\ge 2$, conditioned on $i\in
    \set{S}$.  We will show that this probability is much smaller than
    $1/\alpha k$.  Since each item $h \in M(j) \setminus \{i\}$ is
    chosen independently with probability $\frac{x_h}{\alpha k}$ (even
    given $i\in \set{S}$):
    \begin{equation*}
      \Prob\left[|\set{S}\bigcap\left(M(j)\setminus \{i\}\right)|\ge 2
        ~\big| ~i\in \set{S}\right] \le \frac12 \cdot \left(
        \sum_{h\in M(j)} \frac{x_h}{\alpha k}\right)^2\le
      \frac{\ell^2}{2\alpha^2k^2}
    \end{equation*}
    where the last inequality follows from the fact that
    $$1\ge \sum_{h\in M(j)} s_{hj}\cdot x_h\ge \frac1\ell \sum_{h\in M(j)}
    x_h$$ (recall each item in $M(j)$ has size at least
    $\frac1\ell$). Combining these two cases, we have the desired
    upper bound on $\Prob[E_{ij}\mid i\in \set{S}]$.
  \end{proof}

  \begin{claim}\label{cl:tiny-bnd}
    For any $i\in [n]$, $j\in [m]$ s.t. item $i$ is tiny for
    constraint $j$, $\Prob[E_{ij}\mid i\in \set{S}] \le \frac1{\alpha
      k}\left(1+\frac{2}{\ell}\right)$.
  \end{claim}
  \begin{proof}
    Since $i$ is tiny, if event $E_{ij}$ occurs then the total size
    (in constraint $j$) of items $\set{S}\setminus \{i\}$ is
    greater than $1-\frac1\ell$. So,
    $$ \Prob[E_{ij}\mid i\in \set{S}]
    \le \Prob\left[\sum_{h\in \set{S}\setminus \{i\}} s_{hj} > 1-\frac1\ell\right]
    \le \frac1{\alpha k} \cdot \frac\ell{\ell-1}
    \le \frac1{\alpha k}\left(1+\frac2\ell\right)$$
    where the first inequality follows from the above observation and
    the fact that $\set{S}\setminus \{i\}$ is independent of the event
    $i\in \set{S}$, the second is Markov's inequality, and the last
    uses $\ell\ge 2$.
  \end{proof}

  Thus, for any item $i$ and constraint $j \in N(i)$,
  $\Prob[E_{ij} \mid i \in \set{S} ]
  \le \frac{1}{\alpha k}\max\{(1 + \frac{2}{\ell}),
  (1 + \frac{\ell^2}{2 \alpha k})\}$.
  From the choice of $\ell = (4 \alpha k)^{1/3}$, which makes the
  probability in Claims \ref{cl:med-bnd} and \ref{cl:tiny-bnd} equal,
  we obtain the lemma.
\end{proof}

We now prove the main result of this section
\begin{thm}\label{thm:better-prob}
  For each $i\in [n]$, probability $\Prob[i\in \set{S}'\mid i\in \set{S}] \ge
  \left(1-\frac{1}{\alpha k}\left(1 + (\frac2{\alpha k})^{1/3}\right) \right)^k$.
\end{thm}
\begin{proof}
For any item $i$ and constraint $j \in N(i)$, the conditional event $\left( \neg E_{ij}\mid i\in \set{S}\right)$ is a
\emph{decreasing} function over  the choice of items in set $[n]\setminus \{i\}$. Thus, by the
  FKG inequality~\cite{AS-book}, for any fixed item $i\in[n]$, the probability that no event $(E_{ij}\mid i\in\set{S})$  occurs is:
  $$ \Prob\left[\bigwedge_{j \in N(i)} \neg E_{ij} ~\big|~ i\in \set{S}\right] \ge \prod_{j \in N(i)}
  \Prob[\neg E_{ij} \mid i\in \set{S}]$$

  From Lemma~\ref{lem:prob-eij}, $\Prob[\neg E_{ij}\mid i\in \set{S}] \ge
  1-\frac{1}{\alpha k}\left(1 + (\frac2{\alpha k})^{1/3}\right)$. As
  each item is in at most $k$ constraints, we obtain the theorem.
\end{proof}

Now, by setting $\alpha = 1$,\footnote{Note that this is optimal only
  asymptotically; in the case of $k = 2$, for instance, it is better
  to choose $\alpha \approx 2.8$.} we have $\Prob[i \in \set{S}] =
1/k$, and $\Prob[i \in \set{S'} \mid i \in \set{S}] \ge \frac{1}{e +
  o(1)}$, which immediately implies:

\begin{thm}\label{thm:better-result}
  There is a randomized $(ek+o(k))$-approximation algorithm for \kcs.
\end{thm}

\noindent {\em Remark:} We note that this algorithm can be derandomized using conditional expectation and pessimistic
estimators, since we can compute exactly estimates of the relevant probabilities. Also, using ideas from~\cite{S01} the
algorithm can be implemented in RNC. We defer details to the full version.

\medskip

\noindent {\bf Integrality Gap of LP~(\ref{lp:first}-\ref{lp:last}).} Recall that the LP relaxation for the $k$-set
packing problem has an integrality gap of $k-1+1/k$, as shown by the instance given by the projective plane of order
$k-1$. If we have the same size-matrix and set each capacity to $2-\epsilon$, this directly implies an integrality gap
arbitrarily close to $2 (k-1+1/k)$ for the (weak) LP relaxation for \kcs. This is because the LP can set each $x_i =
(2-\epsilon)/k$ hence obtaining a profit of $(2-\epsilon) (k -1 +1/k)$, while the integral solution can only choose one
item. However, for our stronger LP relaxation~(\ref{lp:first}-\ref{lp:last}) used in this section, this example does
not work and the projective plane instance only implies a gap of $k-1+1/k$ (note that here each item is big in every
constraint that it appears in).

However, using a different instance of \kcs, we show that even the
stronger LP relaxation has an integrality gap at least
$2k-1$. Consider the instance on $n=m=2k-1$ items and constraints
defined as follows. We view the indices $[n]=\{0,1,\cdots,n-1\}$ as
integers modulo $n$. The weights $w_i=1$ for all $i\in[n]$. The sizes
are:
$$s_{ij} := \left\{
\begin{array}{ll} 1& \mbox{ if }i=j\\ \epsilon & \mbox{ if
}j\in\{i+1,\cdots,i+k-1 \mbox{ (mod $n$)}\}\\ 0 & \mbox{ otherwise}
\end{array}\right.,\qquad \forall i,j\in [n].
$$
where $\epsilon>0$ is arbitrarily small, in particular $\epsilon \ll
\frac1{nk}$.

Observe that setting $x_i=1-k \epsilon$ for all $i\in[n]$ is a
feasible fractional solution to the strengthened
LP~(\ref{lp:first}-\ref{lp:last}); each constraint has only one big
item and so the new constraint~\eqref{eq:new-constraint} is
satisfied. Thus the optimal LP value is at least $(1-k \epsilon)\cdot
n\approx n=2k-1$.

On the other hand, we claim that the optimal integral solution can
only choose one item and hence has value 1.  For the sake of
contradiction, suppose that it chooses two items $i,h\in [n]$. Then
there is some constraint $j$ (either $j=i$ or $j=h$) that implies
either $x_i+\epsilon\cdot x_h\le 1$ or $x_h+\epsilon\cdot x_i\le 1$;
in either case constraint $j$ would be violated.

Thus the integrality gap of the LP we consider is at least $2k-1$, for every $k\ge 1$.

\medskip

\noindent {\bf Bad example for possible generalization.} A natural
extension of the \kcs result is to consider PIPs where the
$\ell_1$-norm of each column is upper-bounded by $k$ (when capacities
are all-ones). We observe that unlike \kcs, the LP relaxation for this
generalization has an $\Omega(n)$ integrality gap. The example has
$m=n$; sizes $s_{ii}=1$ for all $i\in [n]$, and $s_{ij}=\frac1n$ for
all $i\ne j$; and all weights one. The $\ell_1$-norm of each column is
at most $2$. Clearly, the optimal integral solution has value one. On
the other hand, picking each column to the extent of $1/2$ is a
feasible LP solution of value $n/2$.

This integrality gap is in sharp contrast to the results on {\em
  discrepancy of sparse matrices}, where the classic Beck-Fiala bound
of $O(k)$ applies also to matrices with entries in $[-1,1]$, just as
well as $\{-1,0,1\}$ entries; here $k$ denotes an upper-bound on the
$\ell_1$-norm of the columns.

\section{Submodular Objective Functions}\label{sec:submod}

We now consider the more general case when the objective we seek to
maximize is an arbitrary {\em monotone submodular function}
$f:2^{[n]}\rightarrow \mathbb{R}_+$. The problem we consider is:
\begin{equation}\label{prob:submod} \max\left\{f(T) ~\big|~ \sum_{i\in
T} s_{ij}\le c_j,~\forall j\in[m];~T\sse [n]\right\}
\end{equation}

As is standard when dealing with submodular functions, we only assume
{\em value-oracle} access to the function: i.e.  the algorithm can
query any subset $T\sse [n]$, and it obtains the function value $f(T)$
in constant time. Again, we let $k$ denote the column-sparseness of
the underlying constraint matrix. Observe that this problem is a
common generalization of maximizing submodular functions over: $k$
partition matroids, and $k$ knapsack constraints.
In this section we obtain an $O(k)$-approximation algorithm for Problem~\eqref{prob:submod}. The algorithm is similar
to that for \kcs (where the objective was additive), and involves the following two steps.
\begin{enumerate}
\item We first solve (approximately) a suitable continuous relaxation
  of~\eqref{prob:submod}. This step follows directly from the
  algorithm of Vondr{\'a}k~\cite{V08}.
\item Then, using the fractional solution, we perform the randomized
  rounding with alteration described in Section~\ref{sec:kcs}.
  Although the algorithm is the same as for additive functions, the
  analysis requires considerably more work. In the process, we also
  establish a new property of submodular functions that generalizes
  {\em fractional subadditivity}~\cite{F06}.
\end{enumerate}

\noindent {\bf Solving the Continuous Relaxation.} The {\em
  extension-by-expectation} (also called the \emph{multi-linear}
extension) of a submodular function $f$ is a continuous function $F:[0,1]^n\rightarrow \mathbb{R}_+$ defined as
follows:
$$F(x) := \sum_{T\sse [n]} \Pi_{i\in T} ~x_i \cdot \Pi_{j\not\in T} ~(1-x_j) \cdot f(T)$$

Note that $F(x) = f(x)$ for $x \in \{0,1\}^n$ and hence $F$ is an extension of $f$. Even though $F$ is a non-linear
function, using the continuous greedy algorithm from Vondr{\'a}k~\cite{V08}, we can obtain a
$\left(1-\frac1e\right)$-approximation algorithm to the following {\em fractional relaxation} of~\eqref{prob:submod}.
\begin{equation} \label{prob:submod-frac}
  \max\left\{F(x) ~\big|~ \sum_{i=1}^n s_{ij}\cdot x_i\le c_j,
    ~\forall j\in[m];~0\le x_i\le 1,~\forall i\in  [n]\right\}
\end{equation}

In order to apply the algorithm from~\cite{V08}, one needs to solve in polynomial time the problem of maximizing a {\em
linear} objective over the constraints $\{ \sum_{i=1}^n s_{ij}\cdot x_i\le c_j,~\forall j\in[m];~0\le x_i\le 1,~\forall
i\in [n]\}$. This is indeed possible since it is a linear program on $n$ variables and $m$ constraints.

\medskip

\noindent {\bf The Rounding Algorithm.} The rounding algorithm is identical to that for \kcs.  Let $x$ denote any
feasible solution to Problem~\eqref{prob:submod-frac}. We apply the rounding algorithm for the additive case (from the
previous section), to first obtain a (possibly infeasible) solution $\set{S}\sse[n]$ and then {\em feasible integral
solution} $\set{S}'\sse[n]$. In the rest of this section, we prove the performance guarantee of this algorithm.

\medskip
\noindent {\bf Fractional Subaddivity.} The following is a useful lemma (see Feige~\cite{F06}) showing that submodular
functions are also fractionally subadditive.
\begin{lemma}[\cite{F06}]\label{lem:subadd}
  Let $\set{U}$ be a set of elements and $\{\set{A}_t \sse
  \set{U}\}$ be a collection of subsets with non-negative weights
  $\{\lambda_t\}$ such that $\sum_{t \mid i \in \set{A}_t}
  \lambda_t \ge 1$ for all elements $i \in \set{U}$. Then, for any
  submodular function $f$, we have $f(\set{U}) \le \sum_{t} \lambda_t
  f(\set{A}_t)$.
\end{lemma}


The above result can be used to show that (the infeasible solution) $\set{S}$ has good profit in expectation.

\begin{lemma}\label{lem:good-S}
  For any $x \in [0,1]^n$ and $0 \le p \le 1$, let set $\set{S}$ be
  constructed by selecting each item $i \in [n]$ independently with
  probability $p \cdot x_i$. Then, $E[f(\set{S})] \ge p F(x)$. In particular, this implies
that our rounding algorithm that forms set $\set{S}$ by independently selecting each element $i\in [n]$ with
probability $x_i/(\alpha k)$ satisfies $E[f(\set{S})] \ge \frac{1}{\alpha k} F(x)$.
\end{lemma}

\begin{proof}
  Consider the following equivalent procedure for constructing
  $\set{S}$: First, construct $\set{S}_0$ by selecting each item $i$
  with probability $x_i$. Then construct $\set{S}$ by retaining each
  element in $\set{S}_0$ independently with probability $p$.

  By definition $E[f(\set{S}_0)] = F(x)$. For any fixed set $T \sse
  [n]$,
consider the outcomes for set $\set{S}$ {\em conditioned} on
  $\set{S}_0=T$; the set $\set{S}\sse \set{S}_0$ is a random subset
  such that $\Prob[i\in \set{S} \mid \set{S}_0=T]=p$ for all $i\in
  T$. Thus by Lemma~\ref{lem:subadd}, we have $E[f(\set{S})\mid
  \set{S}_0=T]\ge p \cdot f(T)$. Hence:

  $$E[f(\set{S})] =
  \sum_{T\sse [n]} \Prob[\set{S}_0=T] \cdot E[f(\set{S})\mid \set{S}_0=T]
  \ge \sum_{T\sse [n]} \Prob[\set{S}_0=T] \cdot p \, f(T)
  = p \, E[f(\set{S}_0)] = p \cdot F(x).$$

 Thus we obtain the lemma.\end{proof}

\medskip

However, the analysis approach in Theorem~\ref{thm:better-prob} does not work.  The problem is that even though
$\set{S}$ (which is chosen by random sampling) has good expected profit, i.e. $E[f(\set{S})] = \Omega(\frac{1}{k})
F(x)$ (from Lemma \ref{lem:good-S} above), it may happen that the alteration step used to obtain $\set{S}'$ from
$\set{S}$ may end up throwing away essentially all the profit. This was not an issue for linear objective functions
since our alteration procedure guarantees that $\Pr[ i\in \set{S'}| i\in \set{S}] = \Omega(1)$ for each $i\in[n]$, and
if $f$ is linear, this implies $E[f(\set{S})] =\Omega(1)\, E[f(\set{S'})]$. However, this property is not enough for
general monotone
submodular functions.  Consider the following:\\

\noindent {\bf Example:} Let set $\set{S} \subseteq [n]$ be drawn from the following distribution:
\begin{itemize}
  \item With probability $1/2n$, $\set{S} = [n]$.
  \item For each $i \in [n]$, $\set{S} = \{i\}$ with
    probability $1/2n$.
  \item With probability $1/2 - 1/2n$, $\set{S} = \emptyset$.
\end{itemize}

Now define $\set{S}' = \set{S}$ if $\set{S} = [n]$, and $\set{S}' = \emptyset$ otherwise. Note that for each $i \in
[n]$, we have $\Pr[i \in \set{S}' \mid i \in \set{S}] = 1/2 =\Omega(1)$.  However, consider the profit with respect to
the ``coverage'' submodular function $f$, where $f(T)=1$ if $T\neq \emptyset$ and is $0$ otherwise.
We have $E[f(\set{S})] =1/2 + 1/2n$, but $E[f(\set{S}')]$ is only $1/2n \ll E[f(\set{S})]$.
\bigskip

{\em Remark:} Note that if $\set{S}'$ itself was chosen randomly from $\set{S}$ such that $\Pr[i \in \set{S}'|
\set{S}=T] = \Omega(1)$ for {\em every $T\sse[n]$ and $i\in T$}, then we would be done by Lemma~\ref{lem:subadd}.
Unfortunately, this is too much to hope for. In our rounding procedure, for any particular choice of $\set{S}$, set
$\set{S'}$ is a fixed subset of $\set{S}$; and there could be (bad) sets $\set{S}$, where after the alteration step we
end up with sets $\set{S}'$ such that $|\set{S}'| \ll |\set{S}|$.

However, it turns out that we can use the following two additional
properties beyond just marginal probabilities to argue that $\set{S'}$
has reasonable profit.  First, the sets $\set{S}$ constructed by our
algorithm are drawn from a product distribution on the items; in
contrast, the example above does not have this property.
  Second, our alteration procedure has the following `monotonicity'
  property: Suppose $T_1 \sse T_2 \sse [n]$, and $i \in \set{S}'$ when
  $\set{S} = T_2$. Then we are guaranteed that $i \in \set{S}'$ when
  $\set{S} = T_1$. (That is, if $\set{S}$ contains additional items,
  it is more likely that $i$ will be discarded by some constraint it
  participates in.) The above example does not satisfy this property either.
%
%
%
%
%
%
%
That these properties suffice is proved in Corollary~\ref{cor:good-S'}.
Roughly speaking, the intuition is that, since $f$ is submodular, the marginal contribution of item $i$ to $\set{S}$ is
largest when $\set{S}$ is ``small'', and this is also the case when $i$ is most likely to be retained for $\set{S}'$.
That is, for every $i\in[n]$, both $\Pr[i \in \set{S}' \mid i \in \set{S}]$ and the marginal contribution of $i$ to
$f(\set{S})$ are \emph{decreasing} functions of $\set{S}$.  To show Corollary~\ref{cor:good-S'} we need the following
generalization of Feige's Subadditivity Lemma.

\begin{thm}\label{thm:subadd}
  Let $[n]$ denote a groundset, $x \in [0,1]^n$, and for each $B\sse [n]$ define $p(B) = \Pi_{i \in B} x_i \cdot \\
  \Pi_{j \notin B} (1-x_j)$. Associated with each $B\sse [n]$, there
  is an arbitrary distribution over subsets of $B$, where each set
  $A\sse B$ has probability $q_B(A)$; so $\sum_{A\sse B} q_B(A)=1$ for
  all $B\sse[n]$. That is, we choose $B$ from a product distribution,
  and then retain a subset $A$ of $B$ by applying a randomized
  alteration.

  \medskip \noindent
  Suppose that the system satisfies the following conditions.

  \vspace{2mm}
  \noindent {\bf Marginal Property:}
  \begin{equation} \label{eqn:submod-hypothesis1}
    \forall i\in[n], \quad \sum_{B\sse[n]}  p(B) \sum_{A\sse B: i \in A} q_B(A)
    \, \, \geq \, \, \beta \cdot \sum_{B\sse[n] : i\in B} p(B).
  \end{equation}
  {\bf Monotonicity:}
  For any two subsets $B\sse B'\sse [n]$ we have,
  \begin{equation} \label{eqn:submod-hypothesis2}
    \forall i \in B, \quad \sum_{A\sse B: i \in A} q_B(A)
      \, \, \geq \, \, \sum_{A'\sse B': i \in A'} q_{B'}(A')
  \end{equation}
  Then, for any monotone submodular function $f$,
  \begin{equation} \label{eqn:submod-dom1}
    \sum_{B\sse [n]} p(B) \, \sum_{A\sse B} q_B(A) \cdot f(A) \,\,
    \geq \,\,  \beta \cdot \sum_{B\sse[n]} p(B) \cdot f(B).
  \end{equation}
\end{thm}

\begin{proof}
  The proof is by induction on $n$, the size of the groundset. The base case of $n = 1$ is
  straightforward.  So suppose $n \geq 2$.  For any subsets $A\sse B\sse [n]$ such
  that $n \in A$, by submodularity we have that $ f(A) \geq f(B) - f(B \setminus \{n\}) + f(A \setminus \{n\}) $.
Applying this,  the left-hand-side of~\eqref{eqn:submod-dom1} is:
\begin{small}
\begin{eqnarray}
&&\sum_{B\sse[n]} p(B) \left( \sum_{A\sse B: n \in A} q_B(A) f(A)\, +
\sum_{A\sse B: n \notin A} q_B(A) f(A) \right) \notag \\
&\ge &  \sum_{B\sse[n]} p(B) \left( \sum_{A\sse B: n \in A} q_B(A) \cdot \bigg( f(B) - f(B \setminus \{n\}) + f(A
\setminus
\{n\})\bigg) + \sum_{A\sse B: n \notin A} q_B(A) f(A) \right) \notag\\
& = & \sum_{B\sse[n]} p(B) \sum_{A\sse B} q_B(A) \cdot f(A\setminus\{n\})  \,\,+ \sum_{B\sse[n]} p(B) \sum_{A\sse B: n
\in A} q_B(A) \cdot \bigg( f(B) - f(B \setminus \{n\})\bigg) \label{eq2}
\end{eqnarray}
\end{small}

Next, we need the following inequality,
\begin{equation} \label{eq:submod-ind}
\sum_{B\sse[n]} \sum_{A\sse B}  p(B) \cdot q_B(A) \cdot f(A\setminus \{n\})  \,\, \geq \,\,  \beta \sum_{B\sse[n]} p(B)
\cdot f(B\setminus \{n\})
\end{equation}
This inequality actually follows by induction, by applying (\ref{eqn:submod-dom1}) to suitably constructed
distributions $p$ and $q$ on subsets of $[n-1]$. We first complete the proof of the theorem
using~\eqref{eq:submod-ind}, and prove~\eqref{eq:submod-ind} later.

We now claim that it suffices to show the following.
  \begin{equation}
    \label{eq3}
    \sum_{B\sse[n]} p(B) \sum_{A\sse B: n \in A} q_B(A) \cdot \big(f(B) - f(B\setminus \{n\}) \big)
    \,\, \geq \,\, \beta \cdot \sum_{B\sse[n]} p(B) \cdot \big( f(B) - f(B\setminus \{n\}) \big).
  \end{equation}
  To see that this suffices, observe that upon
  adding~\eqref{eq:submod-ind} to~\eqref{eq3} we obtain that the right
  hand side of \eqref{eq2} is at least the right hand side of
  \eqref{eqn:submod-dom1}, which will imply the result by
  \eqref{eq2}. We now focus on proving \eqref{eq3}.

\smallskip

\noindent Firstly, note that if $x_n=0$ then~\eqref{eq3} is trivially true. In the following we assume $x_n>0$.

For any set $Y\sse [n-1]$, define the following two functions:
$$g(Y) := f(Y\cup\{n\}) -  f(Y), \quad \mbox{ and }\quad
h(Y) := \sum_{A\sse Y} q_{Y\cup\{n\}} (A\cup\{n\}).$$ Clearly both $g$
and $h$ are non-negative. Note that $g$ is a decreasing function due
to submodularity of $f$.  Moreover function $h$ is also decreasing:
for any $Y\sse Y'\sse[n-1]$,
$$h(Y) = \sum_{A\sse Y}  q_{Y\cup\{n\}} (A\cup\{n\}) \,\, \ge \,\, \sum_{A'\sse Y'}  q_{Y'\cup\{n\}} (A'\cup\{n\}) = h(Y'),$$
where the inequality is by the monotonicity condition with $i=n$, $B=Y\cup\{n\}$ and $B'=Y'\cup\{n\}$.

Consider the product probability space on $2^{[n-1]}$ with marginal probabilities given by $\{x_i\}_{i=1}^{n-1}$. For
any $Y\sse[n-1]$, let $p'(Y) = \Pi_{i \in Y} x_i \cdot \Pi_{j \in [n-1]\setminus Y} (1-x_j)$ denote its probability.
Applying  the FKG inequality~\cite{AS-book} on the {\em decreasing functions} $g$ and $h$, it follows that
  \begin{equation} \label{fkg}
    \sum_{Y\sse[n-1]} p'(Y) \cdot g(Y) \cdot h(Y) \,\, \geq\,\,
    \left( \sum_{Y\sse[n-1]} p'(Y) \, g(Y) \right) \,\cdot \, \left( \sum_{Y\sse[n-1]} p'(Y) \, h(Y) \right).
  \end{equation}

Observe that
\begin{small}  $$ \sum_{Y\sse[n-1]} p'(Y) \cdot h(Y) =
\sum_{Y\sse[n-1]} \frac{p(Y\cup\{n\})}{x_n} \sum_{A\sse Y}
  q_{Y\cup\{n\}} (A\cup\{n\}) = \frac1{x_n} \sum_{B\sse[n]}  p(B) \sum_{A\sse B: n \in A} q_B(A),
$$
\end{small}
which by the Marginal property with $i=n$ is at least $\frac1{x_n}\cdot \beta\, x_n=\beta$. Combining this with
(\ref{fkg}),
\begin{small} \begin{equation}
    \sum_{Y\sse[n-1]} p'(Y) \cdot g(Y) \cdot h(Y) \,\, \geq\,\,
    \beta \, \sum_{Y\sse[n-1]} p'(Y) \cdot g(Y).
  \end{equation}
\end{small}
Using the definitions of $g$ and $h$ (and multiplying both sides by $x_n$), we obtain (\ref{eq3}).

\medskip

\noindent {\bf Proof of Inequality~\eqref{eq:submod-ind}.} We show that this follows by applying Theorem
\ref{thm:subadd} suitably on the groundset $[n-1]$, with marginal probabilities $\{x_i\}_{i=1}^{n-1}$. For each
$C\sse[n-1]$ define $p'(C) = \Pi_{i \in C} x_i \cdot \Pi_{j \in [n-1]\setminus C} (1-x_j)$. Associated with each $C\sse
[n-1]$, let us define the distribution $\{q'_C(A) \mid A\sse C\}$ over subsets of $C$ as follows:
$$ q'_C(A):= x_n\cdot q_{C\cup \{n\}}(A) + x_n\cdot q_{C\cup \{n\}} (A\cup \{n\}) + (1-x_n)\cdot q_C(A),\quad \mbox{ for all } A\sse C\sse [n-1].$$
Note that $\sum_{A: A\sse C \sse [n-1]} q'_C(A) = 1$ for every $C\sse[n-1]$, since
$$\sum_{A: A\sse C} q'_C(A) = x_n \cdot \sum_{A: A \sse C \cup \{n\} } q_{C \cup \{n\}}(A) +  (1-x_n) \cdot \sum_{A: A\sse C} q_C(A)
= x_n + (1-x_n) =1,$$ using the fact that $\sum_{A: A \sse C \cup \{n\} } q_{C \cup \{n\}}(A) = \sum_{A: A \sse C}
q_{C}(A) =1$.

\medskip

\noindent To see that the {\em Marginal property} holds, for any $i\in[n-1]$, we have:
\begin{eqnarray*}
& &\sum_{C\sse[n-1]}  p'(C) \sum_{A\sse C: i \in A} q'_C(A)   \\
& = & \sum_{C\sse[n-1]} p'(C) \sum_{A\sse C: i \in A} \bigg( x_n\cdot q_{C\cup \{n\}}(A) + x_n\cdot q_{C\cup \{n\}} (A\cup \{n\}) + (1-x_n)\cdot q_C(A) \bigg) \\
& = & \sum_{C\sse[n-1]} x_n \, p'(C)  \sum_{A\sse C\cup \{n\}: i \in A} q_{C\cup \{n\}}(A)
+ \sum_{C\sse[n-1]} (1-x_n) \, p'(C) \sum_{A\sse C: i \in A} q_C(A) \\
& = & \sum_{C\sse[n-1]} p(C \cup\{n\}) \sum_{A\sse C \cup \{n\}: i \in A} q_{C\cup \{n\}}(A) + \sum_{C\sse[n-1]} p(C) \sum_{A\sse C: i \in A} q_C(A) \\
& = &  \sum_{B \sse[n]} p(B) \sum_{A\sse B: i \in A} q(A)    \,\,  \geq  \,\, \beta \sum_{B\sse[n] : i\in B} p(B) \,\,
= \,\, \beta \sum_{C\sse[n-1] : i\in C} p'(C).
\end{eqnarray*}
Above, the inequality is by the Marginal property for the original instance on $[n]$.

\medskip

\noindent To show the {\em Monotonicity property} for any subsets $C\sse C'\sse [n-1]$, observe that:
\begin{eqnarray*}
\sum_{A\sse C: i\in A}  q'_C(A) & = & x_n \sum_{A\sse C: i \in A} \bigg( q_{C\cup \{n\}}(A)+ q_{C\cup
\{n\}}(A\cup\{n\}) \bigg)  +
(1-x_n) \sum_{A\sse C: i \in A} q_{C}(A) \\
& = & x_n \sum_{A'\sse C\cup \{n\}: i \in A'} q_{C\cup\{n\}}(A') + (1-x_n) \sum_{A\sse C: i \in A} q_{C}(A) \\
&\ge & x_n \sum_{A'\sse C'\cup \{n\}: i \in A'} q_{C'\cup\{n\}}(A') + (1-x_n) \sum_{A\sse C': i \in A} q_{C'}(A) \\
&=& \sum_{A\sse C': i\in A}  q'_{C'}(A)
\end{eqnarray*}
Again, the inequality is by the monotonicity property on the original instance (on groundset $[n]$) for the pairs
$C\sse C'$ and $C\cup\{n\} \sse C'\cup\{n\}$.

\medskip

\noindent Finally, we can express the left-hand-side of inequality~\eqref{eq:submod-ind} as:
\begin{small}\begin{eqnarray*}
& &\sum_{B\sse[n]}  p(B) \sum_{A\sse B} q_B(A) \,f(A\setminus \{n\})  \\
&=& \sum_{C\sse[n-1]}  \bigg( p(C\cup\{n\}) \sum_{A\sse C\cup\{n\}} q_{C\cup \{n\}}(A) \,f(A\setminus \{n\}) \,\, +
\,\, p(C) \sum_{A\sse C} q_C(A) \,f(A)\bigg)\\
&=& \sum_{C\sse[n-1]}  \left( x_n\cdot p'(C) \sum_{A'\sse C} \big(q_{C\cup\{n\}}(A') + q_{C\cup\{n\}}(A'\cup\{n\})\big)
\,f(A') + (1-x_n)p'(C) \sum_{A'\sse C} q_C(A') \,f(A')\right)\\
&=& \sum_{C\sse[n-1]} p'(C) \sum_{A'\sse C} q'_C(A') \,f(A')\\
&\ge & \beta \sum_{C\sse[n-1]} p'(C) f(C) \quad =\quad \beta \sum_{C\sse[n-1]} \big( p(C) + p(C\cup \{n\}) \big) f(C)
\quad=\quad \beta \sum_{B\sse[n]} p(B) f(B\setminus \{n\}),
\end{eqnarray*}
\end{small}which equals the right-hand-side of~\eqref{eq:submod-ind}. Above, the inequality is by the {\em induction hypothesis} on
the instance on $[n-1]$. This completes the proof of Inequality~\eqref{eq:submod-ind}, and Theorem~\ref{thm:subadd}.
\end{proof}

\noindent {\bf Remark:} It is easy to see that Theorem~\ref{thm:subadd} generalizes Lemma~\ref{lem:subadd}: Let $x_i=1$
for each $i\in [n]$.
The distribution $\left\{ \set{A}_t, \frac{\lambda_t}{\sum_\ell \lambda_\ell}\right\}$ is associated with $B=[n]$. For
all other $B'\subsetneq [n]$, its distribution has $q_{B'}(B')=1$. The monotonicity condition is trivially satisfied.
By the assumption in Lemma~\ref{lem:subadd}, the Marginal property holds with $\beta=1/\sum_\ell \lambda_\ell$. Thus
Theorem~\ref{thm:subadd} applies and yields the conclusion in Lemma~\ref{lem:subadd}.


\begin{cor}\label{cor:good-S'}
Let $\set{S}$ be a random set drawn from a product distribution on $[n]$. Let $\set{S'}$ be another  random set where
for each choice of $\set{S}$, set $\set{S'}$ is an arbitrary subset of $\set{S}$. Suppose that for each $i \in [n]$ the
following hold.
  \begin{itemize}
    \item $\Prob_{\set{S}}[i \in \set{S}' \mid i \in \set{S}] \ge \beta$, and
    \item For all $T_1 \sse T_2$ with $T_1\ni i$, if $i \in \set{S'}$ when $\set{S} =
      T_2$ then $i \in \set{S'}$ when $\set{S} = T_1$.
  \end{itemize}
  Then $E[f(\set{S}')] \ge \beta E[f(\set{S})]$.
\end{cor}
\begin{proof}
This is immediate from Theorem~\ref{thm:subadd}; we simply associate the single set distribution (i.e. $A= \set{S}'$)
for each choice $B$ of $\set{S}$. The two conditions stated above on the construction of $\set{S}'$ imply the Marginal
and Monotonicity properties respectively; and inequality~\eqref{eqn:submod-dom1} translates to $E[f(\set{S}')] \ge
\beta E[f(\set{S})]$.
\end{proof}


We are now ready to prove the performance guarantee of our algorithm. Observe that our rounding algorithm satisfies the
hypothesis of Corollary~\ref{cor:good-S'} with $\beta=\frac1{e+o(1)}$, when parameter $\alpha=1$.
Moreover, by Lemma \ref{lem:good-S}, it follows that $E[f(\set{S})] \geq F(x)/(\alpha k)$.
 Thus,
$$E[f(\set{S'})] \,\, \ge \,\, \frac1{e+o(1)} \, E[f(\set{S})]\,\, \ge \,\, \frac1{ek + o(k)} \cdot F(x),$$
Combined with
the fact that $x$ is an $\frac{e}{e-1}$-approximate solution to the continuous relaxation~\eqref{prob:submod-frac}, we
have proved our main result.

\begin{thm}\label{thm:submodular}
  There is a randomized algorithm for maximizing any monotone
  submodular function over $k$-column sparse packing constraints
  achieving approximation ratio   $\frac{e^2}{e-1}k+o(k)$.
\end{thm}

\section{\kcs Algorithm for general $B$} \label{app:B}

In this section, we obtain substantially better approximation guarantees for \kcs when the capacities are large
relative to the sizes. A useful parameter that measures this is the following (see eg.~\cite{S99}).
$$B:=\min_{i\in[n], j\in[m]} \, \frac{c_j}{s_{ij}}.$$
We consider the \kcs problem as a function of both $k$ and $B$, and obtain an improved approximation ratio of
$O(k^{1/\floor{B}})$; we also give a matching integrality gap (for every $k$ and $B\ge 1$) for the natural LP
relaxation.
Previously, Pritchard \cite{P09} studied \kcs when $B > k$ and obtained a ratio of $(1+k/B)/(1-k/B)$; in contrast, we
obtain improved approximation ratios even when $B = 2$.

\begin{thm}\label{th:B}
  There is a $\left( 4e \cdot \big( (e+o(1))\, \floor{B}\, k\big)^{1/\floor{B}}\right)$-approximation algorithm for
  \kcs, and a  $\left( \frac{4e^2}{e-1} \cdot \big( (e+o(1))\, \floor{B}\, k\big)^{1/\floor{B}}\right)$-approximation
  algorithm for maximizing any monotone submodular function over $k$-column sparse packing constraints.
\end{thm}

It will be convenient to assume that the entries are scaled so that for every constraint $j\in[m]$, $\max_{i\in P(j)}
s_{ij}=1$.  So $B=\min_{j\in[m]} c_j\ge 1$.

Set $\alpha := 4e\cdot (\floor{B}\,k)^{1/\floor{B}}$.  The algorithm first solves the natural LP relaxation for \kcs to
obtain fractional solution $x$. Then it proceeds as follows.

\begin{enumerate}
  \item Sample each item $i\in[n]$ independently with probability
  $x_i/\alpha$. \\Let $\set{S}$ denote the set of chosen items.
  \item Define {\em new sizes} as follows: for every item $i$ and
  constraint $j\in N(i)$, round up $s_{ij}$ to $t_{ij}\in \{2^{-a} \mid a\in \mathbb{Z}_+\}$, the next
  larger power of $2$.
\item For any item $i$ and constraint $j \in N(i)$, let $E_{ij}$
  denote the event that the items $\{i'\in \set{S} \mid t_{i'j}\ge
  t_{ij}\}$ have total $t$-size (in constraint $j$) exceeding one. Mark
  $i$ for deletion if $E_{ij}$ occurs for any $j \in N(i)$.

\item Return set $\set{S}'\sse \set{S}$ consisting of all items $i\in
  \set{S}$ not marked for deletion.
\end{enumerate}

Note the differences from the algorithm in Section~\ref{sec:kcs}: the scaling factor for randomized rounding is
smaller, and the alteration step is more intricate (it uses slightly modified sizes). It is clear that $\set{S}'$ is a
feasible solution with probability one, since the original $s$-sizes are at most the new $t$-sizes.\medskip

The approximation guarantee is proved using the following theorem.
\begin{thm}\label{th:B-prob}
  For each $i\in [n]$, probability $\Prob[i\in \set{S}'\mid i\in
  \set{S}] \ge \left(  1- \frac1{k\,\floor{B}} \right)^k$.
\end{thm}
\begin{proof}
  Fix any $i\in[n]$ and $j\in N(i)$. Recall that $E_{ij}$ is the event
  that items $\{i'\in \set{S}\mid t_{i'j}\ge t_{ij}\}$ have total
  $t$-size (in constraint $j$) greater than $c_j$.

\medskip



  %
We first bound $\Pr[E_{ij}\mid i\in \set{S}]$.  Let $t_{ij}=2^{-\ell}$, where $\ell\in\mathbb{N}$. Observe that all
  the $t$-sizes that are at least $2^{-\ell}$ are actually integral
  multiples of $2^{-\ell}$ (since they are all powers of two). Let
  $\is_{ij} = \{i'\in [n] \mid t_{i'j}\ge t_{ij}\} \setminus \{i\}$,
  and $Y_{ij}:=\sum_{i'\in \is_{ij}} t_{i'j}\cdot \II_{i'\in \set{S}}$
  where $\II_{i'\in \set{S}}$ are indicator random variables. The
  previous observation implies that $Y_{ij}$ is always an integral
  multiple of $2^{-\ell}$. Note that
  \begin{eqnarray*}
    \Prob[E_{ij}\mid i\in \set{S}] =
    \Prob\left[Y_{ij}>c_j-2^{-\ell}\mid i\in \set{S}\right] &\le&
    \Pr\left[Y_{ij}>\lfloor c_j\rfloor -2^{-\ell}\mid i\in \set{S}\right]\\
    &= &\Prob\left[Y_{ij}\ge \lfloor c_j\rfloor \mid i\in \set{S}\right],
  \end{eqnarray*}
  where the last equality uses the fact that $Y_{ij}$ is always a
  multiple of $2^{-\ell}$. Since each item is included into $\set{S}$
  independently, we also have $\Prob[Y_{ij}\ge \lfloor c_j\rfloor \mid
  i\in \set{S}]=\Prob[Y_{ij}\ge \lfloor c_j\rfloor]$. Now $Y_{ij}$ is
  the sum of independent $[0,1]$ random variables with mean:

  $$E[Y_{ij}] = \sum_{i'\in \is_{ij}} t_{i'j}\cdot \Pr[i'\in \set{S}]
  \le \sum_{i'=1}^nt_{i'j}\cdot \frac{x_{i'}}\alpha
  \le \frac2\alpha\sum_{i'=1}^n s_{i'j}\cdot x_{i'}
  \le \frac2\alpha c_j.$$

  Choose $\delta$ such that $(\delta+1)\cdot E[Y_{ij}]={\lfloor c_j
    \rfloor}$, i.e. (using $c_j\ge 1$),
  $$\delta+1 = \frac{\lfloor c_j \rfloor}{E[Y_{ij}]}
  \ge \frac{\alpha\,\lfloor c_j \rfloor}{2\cdot c_j}\ge \frac\alpha4.$$

  Now using Chernoff Bound~\cite{MR-book}, we have:
  $$\Pr[Y_{ij}\ge \lfloor c_j \rfloor]
  = \Pr\left[Y_{ij}\ge (1+\delta) \cdot E[Y_{ij}]\right]
  \le \left(\frac{e}{\delta+1}\right)^{\lfloor c_j\rfloor}
  \le \left(\frac{4\, e}\alpha\right)^{\lfloor c_j\rfloor}
  \le \left(\frac{4\, e}\alpha\right)^{\lfloor B\rfloor}.$$

  The last inequality uses the fact that $c_j\ge B$. Finally, by the choice of
  $\alpha= 4e\cdot (\floor{B}\,k)^{1/\floor{B}}$,

\begin{equation}\label{eq:B-eij}
\Pr[E_{ij}\mid i\in \set{S}] \,\le  \, \Pr[Y_{ij}\ge \lfloor c_j \rfloor] \,\le \, \frac1{k\,\floor{B}}.
\end{equation}

\medskip

As in the proof of Theorem~\ref{thm:better-prob}, for any fixed item $i\in [n]$, the conditional events $\{ E_{ij} \mid
i\in \set{S} \}_{j\in N(i)}$ are positively correlated. Thus using~\eqref{eq:B-eij} and the FKG
inequality~\cite{AS-book},
$$\Pr[i\in \set{S'}\mid i\in \set{S} ] = \Pr\left[\bigwedge_{j \in N(i)} \neg E_{ij}\mid i\in \set{S}\right] \ge \prod_{j \in N(i)}
  \Prob[\neg E_{ij}\mid i\in \set{S}] \ge \left(  1- \frac1{k\,\floor{B}} \right)^k$$
This completes the proof of the theorem.
\end{proof}

As a function of $k$, we obtain that $\Pr[i\in\set{S'} \mid i\in\set{S}]\ge \big( e+ o(1)\big)^{-1/\floor{B}}$. Since
$\Pr[i\in\set{S}]=x_i/\alpha$, we obtain the first part of Theorem~\ref{th:B}.

This algorithm can also be used for maximizing monotone {\em submodular} functions over such packing constraints
(parameterized by $k$ and $B$). Again we would first (approximately) solve the continuous relaxation using~\cite{V08},
and perform the above randomized rounding and alteration. Corollary~\ref{cor:good-S'} can be used with
Theorem~\ref{th:B-prob} to obtain a  $\left( \frac{4e^2}{e-1} \cdot \big( (e+o(1))\, \floor{B}\,
k\big)^{1/\floor{B}}\right)$-approximation algorithm.

This completes the proof of Theorem~\ref{th:B}.

\subsection{Integrality Gap for General $B$}
We show that the natural LP relaxation for \kcs has an $\Omega(k^{1/\lfloor B\rfloor})$ integrality gap for every $B\ge
1$, matching the above approximation ratio up to constant factors.

For any $B\ge 1$, let $t:=\lfloor B\rfloor$. We construct an instance of \kcs with $n$ columns and $m={n \choose t+1}$
constraints. For all $i\in[n]$, weight $w_i=1$.

For every ($t+1$)-subset $C\sse [n]$, there is a constraint $j(C)$ involving the variables in $C$: set $s_{i,j(C)}=1$
for all $i\in C$, and $s_{i,j(C)}=0$ for $i\not\in C$. For each constraint $j \in [m]$, the capacity $c_j=B$. Note that
the column sparsity $k={n-1 \choose
  t}\le(ne/t)^t$.

\bigskip
 Setting $x_i=\frac12$ for all $i\in[n]$ is a feasible
fractional solution. Indeed, each constraint is occupied to extent $\frac{t+1}2\le \frac{B+1}2\le B$ (since $B\ge 1$).
Thus the optimal LP value is at least $\frac{n}2$.

On the other hand, the optimal integral solution has value at most $t$. Suppose for contradiction that the solution
contains some $t+1$ items, indexed by $C\sse [n]$. Then consider the constraint $j(C)$, which is occupied to extent
$t+1=\lfloor B\rfloor +1 >B$, this contradicts the feasibility of the solution! Thus the integral optimum is $t$, and
the integrality gap for this instance is at least $\frac{n}{2t}\ge \frac1{2e}\,k^{1/\lfloor B\rfloor}$.

\paragraph{Acknowledgements:} NK thanks Chandra Chekuri and Alina Ene
for detailed discussions on \kcs.
We also thank Deeparnab Chakarabarty and David Pritchard for discussions and sharing a copy of \cite{CP09}.  We thank
Jan Vondr{\'a}k and Chandra Chekuri for pointing out an error in the original proof of Theorem~\ref{thm:submodular},
which prompted us to prove Theorem~\ref{thm:subadd}. Our thanks also to the IPCO referees for their helpful
suggestions.

\end{document}